\documentstyle[preprint,eqsecnum,aps]{revtex}
\tightenlines
\begin{document}
\title{Casimir Energy in the Axial Gauge}
\author{Giampiero Esposito,$^{1,2}$
\thanks{Electronic address: giampiero.esposito@na.infn.it}
Alexander Yu. Kamenshchik$^{3,4}$
\thanks{Electronic address: landau@icil64.cilea.it}
and Klaus Kirsten$^{5}$
\thanks{Electronic address: klaus@a13.ph.man.ac.uk}}
\address{${ }^{1}$Istituto Nazionale di 
Fisica Nucleare, Sezione di Napoli,\\
Complesso Universitario di Monte S. Angelo, Via Cintia,
Edificio N', 80126 Napoli, Italy\\
${ }^{2}$Dipartimento di Scienze Fisiche,
Complesso Universitario di Monte S. Angelo,\\
Via Cintia, Edificio N', 80126 Napoli, Italy\\
${ }^{3}$L.D. Landau Institute for Theoretical Physics,
Russian Academy of Sciences,\\
Kosygina Str. 2, Moscow 117334, Russia\\
${ }^{4}$Landau Network--Centro Volta,
Villa Olmo, Via Cantoni 1, 22100 Como, Italy\\
${ }^{5}$The University of Manchester, Department of Physics
and Astronomy,\\
Theory Group, Schuster Laboratory,\\
Oxford Road, Manchester M13 9PL, England}
\maketitle
\begin{abstract}
The zero-point energy of a conducting spherical shell is
studied by imposing the axial gauge via path-integral methods,
with boundary conditions on the electromagnetic potential
and ghost fields. The coupled modes are then found to be the
temporal and longitudinal modes for the Maxwell field. The
resulting system can be decoupled by studying a fourth-order
differential equation with boundary conditions on longitudinal
modes and their second derivatives. The exact solution of such
equation is found by using a Green-function method, and is
obtained from Bessel functions and definite integrals involving
Bessel functions. Complete agreement with a previous path-integral
analysis in the Lorenz gauge, and with Boyer's value, is
proved in detail.
\end{abstract}
\section{Introduction}

In recent work by the authors [1] the zero-point energy of a
perfectly conducting spherical shell, a problem first investigated
by Boyer [2], has been studied from the point of view of
path-integral quantization, with the associated boundary conditions
on modes for the potential and the ghost. Since there are good 
reasons for regarding path integrals as a basic tool in the
quantization of gauge fields [3], such a re-derivation of a quantum
effect is a lot more than a useful exercise in field theory. In
particular, one may hope to be able to use similar investigations
to prove the gauge independence of path-integral calculations on
manifolds with boundary, and also to make predictions which can be
tested against observations [4,5]. The key elements of the analysis
in Ref. [1] that we need are as follows. 
\vskip 0.3cm
\noindent
(i) The perfect-conductor boundary conditions for a spherical shell,
according to which tangential components of the electric field should
vanish at the boundary $\partial M$, are satisfied if
\begin{equation}
[A_{t}]_{\partial M}=0,
\label{(1.1)}
\end{equation}
\begin{equation}
[A_{\theta}]_{\partial M}=0,
\label{(1.2)}
\end{equation}
\begin{equation}
[A_{\varphi}]_{\partial M}=0.
\label{(1.3)}
\end{equation}
With our notation, $A_{t}$ is the temporal component of the 
electromagnetic potential $A_{\mu}$, while $A_{\theta}$ and
$A_{\varphi}$ are its tangential components. In the classical
theory, the boundary conditions (1.1)--(1.3) are gauge-invariant
if and only if the gauge function $\varepsilon$ occurring in the
gauge transformations 
\begin{equation}
{ }^{\varepsilon}A_{\mu} \equiv A_{\mu}+\nabla_{\mu}\varepsilon
\label{(1.4)}
\end{equation}
vanishes at $\partial M$:
\begin{equation}
[\varepsilon]_{\partial M}=0.
\label{(1.5)}
\end{equation}
In the quantum theory, Eq. (1.5) is a shorthand notation for the
vanishing at the boundary of two {\it independent} ghost fields
[1,6]. At this stage, the only boundary condition whose preservation
under gauge transformations (1.4) is again guaranteed by Eq. (1.5)
is the vanishing of the gauge-averaging functional 
at the boundary: 
\begin{equation}
[\Phi(A)]_{\partial M}=0.
\label{(1.6)}
\end{equation}
The latter is a map which associates to a one-form $A_{\mu}dx^{\mu}$ a
real number, and reflects the freedom of choosing supplementary
conditions in the classical theory. The square of $\Phi(A)$, divided
by $2\alpha$ (see below), should be added to the Maxwell Lagrangian
to obtain an invertible operator on perturbations of $A_{\mu}$ in the
quantum theory [6].
\vskip 0.3cm
\noindent
(ii) If $\Phi(A)$ is chosen to be of the axial type:
$\Phi(A)=N^{\mu}A_{\mu}$, which is quite relevant for the quantization
program in noncovariant gauges [7], the potential is found to obey
the equation
\begin{equation}
P_{\mu}^{\; \nu} \; A_{\nu}=0,
\label{(1.7)}
\end{equation}
where $P_{\mu}^{\; \nu}$ is the second-order operator
\begin{equation}
P_{\mu}^{\; \nu}=-\delta_{\mu}^{\; \nu}\Box
+\nabla_{\mu}\nabla^{\nu}+{1\over \alpha}N_{\mu}N^{\nu},
\label{(1.8)}
\end{equation}
$\alpha$ being a gauge parameter with dimension length squared.
Moreover, $A_{\nu}$ obeys the equations
\begin{equation}
\nabla_{\mu}(P_{\mu}^{\; \nu}A_{\nu})=0,
\label{(1.9)}
\end{equation}
\begin{equation}
N^{\mu}P_{\mu}^{\; \nu}A_{\nu}=0.
\label{(1.10)}
\end{equation}
The boundary condition (1.6) and the differential equation (1.9)
imply that the normal component of the potential, i.e., 
$A_{r} \equiv {\vec r} \cdot {\vec A}$, vanishes everywhere 
inside the spherical shell, including the boundary [1]:
\begin{equation}
A_{r}(t,r,\theta,\varphi)=0 \; \; \forall r \in [0,R].
\label{(1.11)}
\end{equation}
Moreover, the axial gauge-averaging functional leads to the ghost
operator $Q=-{\partial \over \partial r}$, and hence the boundary
condition (1.5) implies that the ghost field vanishes everywhere 
as well [1]:
\begin{equation}
\varepsilon(t,r,\theta,\varphi)=0 \; \; \forall r \in [0,R].
\label{(1.12)}
\end{equation}
\vskip 0.3cm
\noindent
(iii) One is then left with the temporal component $A_{t}$,
expanded in harmonics on the two-sphere [1]:
\begin{equation}
A_{t}(t,r,\theta,\varphi)=\sum_{l=0}^{\infty}\sum_{m=-l}^{l}
a_{l}(r)Y_{lm}(\theta,\varphi)e^{i \omega t},
\label{(1.13)}
\end{equation}
and tangential components (here $k$ stands for $\theta$ and
$\varphi$)
\begin{equation}
A_{k}(t,r,\theta,\varphi)=\sum_{l=1}^{\infty}\sum_{m=-l}^{l}
\Bigr[c_{l}(r)\partial_{k}Y_{lm}(\theta,\varphi)
+T_{l}(r)\varepsilon_{kp}\partial^{p}Y_{lm}(\theta,\varphi)
\Bigr]e^{i \omega t}.
\label{(1.14)}
\end{equation}
Of course, $c_{l}$ and $T_{l}$ are longitudinal and transverse 
modes, respectively. On setting $\omega=iM$ [1], the latter are
found to obey the eigenvalue equation 
\begin{equation}
\left[{d^{2}\over dr^{2}}-{l(l+1)\over r^{2}}\right]T_{l}
=M^{2}T_{l},
\label{(1.15)}
\end{equation}
whose regular solution reads
\begin{equation}
T_{l}(r)=\sqrt{\pi /2} \sqrt{r}I_{l+{1\over 2}}(Mr) \; \; 
\forall r \in [0,R].
\label{(1.16)}
\end{equation}
\vskip 0.3cm
Section II studies the coupled equations for temporal and
longitudinal modes, with the associated fourth-order equation
for $c_{l}$ modes only. Section III finds the explicit form
of $c_{l}$ modes by means of a Green-function method, and proves
agreement with the Casimir energy found by Boyer. Concluding 
remarks are presented in Sec. IV.

\section{Fourth-order equation for longitudinal modes}

In the axial gauge on a spherical shell, the temporal and
longitudinal modes are known to obey, from the work in 
Ref. [1], the coupled system
\begin{equation}
\left[{d^{2}\over dr^{2}}+{2\over r}{d\over dr}
-{l(l+1)\over r^{2}}\right]a_{l}-{Ml(l+1)\over r^{2}}
c_{l}=0,
\label{(2.1)}
\end{equation}
\begin{equation}
\left[{d^{2}\over dr^{2}}-M^{2}\right]c_{l}-Ma_{l}=0.
\label{(2.2)}
\end{equation}
Although this system cannot be decoupled to find second-order
equations with Bessel-type solutions [1], it can however be easily
decoupled if one expresses $a_{l}$ from Eq. (2.2) and inserts the
result into Eq. (2.1). This leads to the following fourth-order
equation for $c_{l}$:
\begin{equation}
\left[{d^{4}\over dr^{4}}+{2\over r}{d^{3}\over dr^{3}}
-\left(M^{2}+{l(l+1)\over r^{2}}\right){d^{2}\over dr^{2}}
-{2\over r}M^{2}{d\over dr}\right]c_{l}(Mr)=0,
\label{(2.3)}
\end{equation}
while $a_{l}$ is eventually obtained as
\begin{equation}
a_{l}(Mr)={1\over M}\left[{d^{2}\over dr^{2}}-M^{2}\right]
c_{l}(Mr).
\label{(2.4)}
\end{equation}
These equations are dimensionally correct, bearing in mind that, 
from the rule for taking derivatives of composite functions,
$$
{dc_{l}\over dr}=Mc_{l}'(Mr)
$$
and so on, where each prime denotes the derivative evaluated
at $y \equiv Mr$. The variable $y$ is dimensionless, and makes it
possible to study Eq. (2.3) in a form more convenient for the
following calculations, i.e.,
\begin{equation}
\left[{d^{4}\over dy^{4}}+{2\over y}{d^{3}\over dy^{3}}
-\left(1+{l(l+1)\over y^{2}}\right){d^{2}\over dy^{2}}
-{2\over y}{d\over dy}\right]c_{l}(y)=0.
\label{(2.5)}
\end{equation}
Recall now from the Introduction that, at $r=R$, both $a_{l}$
and $c_{l}$ should vanish. By virtue of (2.4), the second derivative
of $c_{l}$ should then vanish as well on the boundary. Moreover,
by regularity at the origin, $a_{l}$ and $c_{l}$ can be set to zero
at $r=0$, which implies that also ${d^{2}c_{l}\over dr^{2}}$ vanishes
at $r=0$. The full set of boundary conditions for Eq. (2.5) can be
therefore taken to be
\begin{mathletters}
\begin{equation}
\left . c_{l} \right |_{y=0}=0,
\label{(2.6a)}
\end{equation}
\begin{equation}
\left . c_{l} \right |_{y=MR}=0,
\label{(2.6b)}
\end{equation}
\end{mathletters}
\begin{mathletters}
\begin{equation}
\left . {d^{2}c_{l}\over dy^{2}} 
\right |_{y=0}=0,
\label{(2.7a)}
\end{equation}
\begin{equation}
\left . {d^{2}c_{l}\over dy^{2}} 
\right |_{y=MR}=0.
\label{(2.7b)}
\end{equation}
\end{mathletters}
At a deeper mathematical level, the boundary conditions
(2.6a), (2.6b), (2.7a) and (2.7b) can be proved to determine a 
domain of self-adjoint extension of a fourth-order differential
operator in one dimension on the closed interval $[0,R]$ (see
sections $2$ and $3$ of Ref. [8]).

\section{Exact solution}

We are interested in solving the fourth-order differential 
equation (2.5) for $c_l$. Since the unknown function $c_{l}$
never occurs undifferentiated therein, a natural
starting point is to put 
\begin{equation}
F_{l}(y) \equiv c_{l} ' (y),  
\label{(3.1)}
\end{equation}
from which the differential equation for $F_{l}(y)$ reads 
\begin{equation}
\left[{d^{3}\over dy^{3}}+{2\over y}{d^{2}\over dy^{2}}
-\left(1+{l(l+1)\over y^{2}}\right){d\over dy}
-{2\over y}\right]F_{l}(y)=0.
\label{(3.2)}
\end{equation}
An obvious solution is $F_{l}(y) =0$ and hence
$c_{l}(y)={\rm const}$. An elegant way to 
proceed is to point out that (\ref{(3.2)}) can be rewritten as 
\begin{equation}
{1\over y^{2}}{d\over dy}\left[y^{2}\left({d^{2}\over dy^{2}}
-\left(1+{l(l+1)\over y^{2}}\right)\right)F_{l}(y)\right]=0.
\label{(3.3)}
\end{equation}
This shows that solutions of Eq. (\ref{(3.2)}) 
can be found as solutions of 
\begin{equation}
\left[{d^{2}\over dy^{2}}-\left(1+{l(l+1)\over y^{2}}\right)
\right]F_{l}(y)={C\over y^{2}},
\label{(3.4)}
\end{equation}
where $C$ is a constant. The general solution of Eq. (\ref{(3.4)}) 
has the form
\begin{equation}
F_{l}(y) = C F_{sp} (y) +d_{1} F_{1} (y) +d_{2} F_{2} (y),
\label{(3.5)} 
\end{equation}
with $F_{sp}(y)$ being a special solution of Eq.
(\ref{(3.4)}) with $C=1$, while $F_{1}$ and $F_{2}$ are the
linearly independent integrals of the homogeneous equation
\begin{equation}
\left[{d^{2}\over dy^{2}}-\left(1+{l(l+1)\over y^{2}}\right)
\right]{\cal F}(y)=0.
\label{(3.6)}
\end{equation}
They read, with our conventions, 
\begin{equation}
F_{1} (y) = \sqrt{\pi / 2} \sqrt{y} I_{l+1/2}(y) ,
\label{(3.7)}
\end{equation}
\begin{equation}
F_{2} (y) = \sqrt{\pi / 2} \sqrt{y} I_{-l-1/2}(y).
\label{(3.8)}
\end{equation} 
To find the solution $F_{sp}(y)$ of Eq. (\ref{(3.4)}) with $C=1$ for all
values of $l$, a Green-function approach is
convenient. On defining
\begin{equation}
L \equiv {d^{2}\over dy^{2}}-\left(1+{l(l+1)\over y^{2}}\right),
\label{(3.9)}
\end{equation}
\begin{equation}
f(y) \equiv {1\over y^{2}},
\label{(3.10)}
\end{equation}
we have first to find the Green function $G(y,\xi)$ which, by
definition [9], solves the equation (hereafter $b \equiv MR$)
\begin{equation}
LG=0 \; \; {\rm for} \; \; y \in ]0,\xi[ \; \; {\rm and} \; \;
y \in ]\xi,b[,
\label{(3.11)}
\end{equation}
the boundary conditions (see (\ref{(2.7a)}), (\ref{(2.7b)}) and
(\ref{(3.1)}))
\begin{equation}
G'(0,\xi)=0,
\label{(3.12)}
\end{equation}
\begin{equation}
G'(b,\xi)=0,
\label{(3.13)}
\end{equation}
the continuity condition
\begin{equation}
\lim_{y \to \xi^{+}}G(y,\xi)=\lim_{y \to \xi^{-}}G(y,\xi),
\label{(3.14)}
\end{equation}
and the jump condition
\begin{equation}
\lim_{y \to \xi^{+}}{\partial G \over \partial y}
-\lim_{y \to \xi^{-}}{\partial G \over \partial y}=1.
\label{(3.15)}
\end{equation}
The general theory of boundary-value problems for second-order 
equations [9] tells us that, when the operator $L$ is studied
with the unmixed boundary conditions of our problem, i.e.
\begin{equation}
F_{l}'(0)=0,
\label{(3.16)}
\end{equation}
\begin{equation}
F_{l}'(b)=0,
\label{(3.17)}
\end{equation}
the construction of $G(y,\xi)$ involves a nontrivial solution
$u_{1}(y)$ of the homogeneous equation $Lu=0$ satisfying
$u'(0)=0$, and a nontrivial solution $u_{2}(y)$ of $Lu=0$ 
satisfying $u'(b)=0$. More precisely, by virtue of
(\ref{(3.11)})--(\ref{(3.13)}) one finds
\begin{equation}
G(y,\xi)=A(\xi)u_{1}(y) \; \; {\rm if} \; \; 
y \in ]0,\xi[,
\label{(3.18)}
\end{equation}
\begin{equation}
G(y,\xi)=B(\xi)u_{2}(y) \; \; {\rm if} \; \;
y \in ]\xi,b[,
\label{(3.19)}
\end{equation}
where $u_{1}$ and $u_{2}$ are linearly independent. By virtue 
of Eqs. (\ref{(3.14)}) and (\ref{(3.15)}),
$A(\xi)$ and $B(\xi)$ are obtained by solving the inhomogeneous
system
\begin{equation}
A(\xi)u_{1}(\xi)-B(\xi)u_{2}(\xi)=0,
\label{(3.20)}
\end{equation}
\begin{equation}
B(\xi)u_{2}'(\xi)-A(\xi)u_{1}'(\xi)=1,
\label{(3.21)}
\end{equation}
which yields
\begin{equation}
A(\xi)={u_{2}(\xi)\over W(u_{1},u_{2};\xi)},
\label{(3.22)}
\end{equation}
\begin{equation}
B(\xi)={u_{1}(\xi)\over W(u_{1},u_{2};\xi)},
\label{(3.23)}
\end{equation}
where $W$ is the Wronskian of $u_{1}$ and $u_{2}$. For the
operator $L$, which is formally self-adjoint, the Wronskian
reads [9]
\begin{equation}
W=\gamma ,
\label{(3.24)}
\end{equation}
$\gamma$ being a constant. Thus, on defining as usual
$y_{<} \equiv {\rm min}(y,\xi), y_{>} \equiv {\rm max}(y,\xi)$,
one finds a very simple formula for the Green function, i.e. [9]
\begin{equation}
G(y,\xi)={1\over \gamma}u_{1}(y_{<}) u_{2}(y_{>}).
\label{(3.25)}
\end{equation}
The particular solution $F_{sp}$ of the
inhomogeneous boundary-value problem given by the equation
\begin{equation}
LF(y)=f(y) \; \; y \in ]0,b[,
\label{(3.26)}
\end{equation}
with boundary conditions (\ref{(3.16)}) and (\ref{(3.17)}) written for
$F_{sp}$:
\begin{equation}
F_{sp}'(0)=F_{sp}'(b)=0,
\label{(3.27)}
\end{equation}
is then given by the integral [9]
\begin{equation}
F_{sp}(y)=\int_{0}^{b}G(y,\xi)f(\xi)d\xi .
\label{(3.28)}
\end{equation}
We may now choose ($a_{1}$ and $a_{2}$ being some parameters)
\begin{equation}
u_{1}(y)=\sqrt{\pi /2} \sqrt{y}I_{l+1/2}(y),
\label{(3.29)}
\end{equation}
\begin{equation}
u_{2}(y)=\sqrt{\pi /2} \sqrt{y}\Bigr[a_{1}I_{l+1/2}(y)+a_{2}I_{-l-1/2}(y)\Bigr],
\label{(3.30)}
\end{equation}
bearing in mind that $u_{1}$ and $u_{2}$ should be linearly independent
and should satisfy Neumann boundary conditions at $0$ and at $b$, respectively,
which gives a relation between $a_{1}$ and $a_{2}$.
However, what is truly essential for us is that the general theory of
one-dimensional boundary-value problems ensures that an $F_{sp}$ exists
satisfying the conditions (3.27).  
The general solution (3.5) of Eq. (3.4) can be
therefore written in the form (see (3.7) and (3.8))
\begin{equation}
F_{l}(y)=C F_{sp}(y)+d_{1}F_{1}(y)+d_{2}F_{2}(y). 
\label{(3.31)}
\end{equation}
Regularity at the origin (see (2.6a)) implies that $d_{2}=0$. The vanishing
of $c_{l}$ at the boundary (see (2.6b)) fixes the relation between $C$ 
and $d_{1}$, i.e.
\begin{equation}
C=-d_{1}{\int_{0}^{b}dz \; F_{1}(z) \over \int_{0}^{b} dz \; F_{sp}(z)}.
\label{(3.32)}
\end{equation}
Eventually, the vanishing of $c_{l}''$ at the boundary (see (2.7b))
leads to 
\begin{equation}
I_{l+1/2}'(b)+{1\over 2b}I_{l+1/2}(b)=0,
\label{(3.33)}
\end{equation}
because, by construction, $C F_{sp}$ is a particular solution of Eq. (3.4)
whose first derivative vanishes at the boundary.

Equation (3.33), jointly with the vanishing at the boundary of the transverse
modes given in Eq. (1.16), yields the same set of eigenvalue conditions,
with the same degeneracies, found for the interior problem in the Lorenz
gauge in Ref. [1]. Thus, complete agreement with Boyer's value for the
Casimir energy is recovered (the exterior problem can be studied on replacing
$I_{l+1/2}$ by $K_{l+1/2}$, without any difficulty).

\section{Concluding remarks}

Although noncovariant gauges break relativistic covariance, they make
it possible to decouple Faddeev--Popov ghosts [10] from the gauge
field. Thus, ghost diagrams do not contribute to cross-sections and
need not be evaluated, 
and this property has been regarded as the main advantage of
noncovariant gauges [7]. In the case of Casimir energies, 
the ghost field is forced to vanish everywhere by virtue
of the boundary conditions appropriate for the axial gauge [1], and the 
analysis of temporal and longitudinal modes, although rather involved,
has been here proved to lead to the same Casimir energy [2] for the interior
problem found in the Lorenz gauge [1]. The particular solution of the
inhomogeneous equation (3.4) plays a nontrivial role in ensuring that,
despite some technicalities, the resulting Casimir energy is the same
as in the Lorenz gauge, hence proving explicitly the equivalence of a
covariant and a non-covariant gauge for a conducting spherical shell.

A better theoretical understanding of gauge independence in quantum field
theory has been therefore gained, after the encouraging experimental progress
of recent years in the measurement of Casimir forces [4,5,11,12].
 
\acknowledgments

The work of G.E. has been partially supported by PRIN97
``Sintesi.'' K.K. thanks Stuart Dowker for interesting discussions. The work
of K.K. has been supported by EPSRC, grant no GR/M08714.

\end{document}